\documentclass[runningheads]{llncs}
\usepackage{graphicx}

\usepackage{float}
\usepackage{amsmath,amssymb,amsfonts}

\usepackage{graphicx}
\usepackage{cleveref}
\usepackage{textcomp}
\usepackage{xcolor}
\usepackage{import}
\usepackage{booktabs}
\usepackage{tabularx}
\usepackage{float}
\usepackage{makecell}
 \usepackage{multirow}
 \usepackage{colortbl}
\usepackage{graphicx}
\usepackage{textcomp}
\usepackage{xcolor}
\usepackage{graphicx}
\usepackage{graphics}
\usepackage{adjustbox}
\usepackage{colortbl}
\usepackage{color}
\usepackage{amssymb,amsfonts}
\usepackage{algpseudocode}
\usepackage[ruled,linesnumbered,lined,boxed,commentsnumbered]{algorithm2e}
\usepackage{graphicx}
\usepackage{colortbl}
\usepackage{lipsum}%
\usepackage{textcomp}
\usepackage{xcolor}
\usepackage{import}
\usepackage{amsmath, multicol, multirow, booktabs, microtype, balance}
\usepackage{subcaption}
\usepackage{float}

\usepackage{mwe} 

\usepackage{booktabs}
\usepackage{siunitx}

\usepackage{colortbl}
\usepackage{colortbl}

\usepackage{colortbl}

\usepackage{multirow}
\usepackage{colortbl}

\usepackage{rotating}
\usepackage{multirow}
\usepackage{colortbl}

\usepackage{multirow}
\usepackage{colortbl}

\usepackage{makecell}
\usepackage{multirow}
\usepackage{colortbl}
\usepackage{multirow}
\usepackage{color}
\usepackage{colortbl}
\usepackage{float}
\begin{document}
\title{Examining the behaviour of state-of-the-art convolutional neural networks for brain tumor detection with and without transfer learning}
%
\author{Md. Atik Ahamed \inst{1}\orcidID{0000-0002-7746-4247} \and
Rabeya Tus Sadia\inst{2}}
%
\institute{Green University of Bangladesh, Dhaka, Bangladesh \\
\email{atikahamed@cse.green.edu.bd}\\
\and
Green University of Bangladesh, Dhaka, Bangladesh\\
\email{rabeya@cse.green.edu.bd}}
\maketitle              
\begin{abstract}
Distinguishing normal from malignant and determining the tumor type are critical components of brain tumor diagnosis. Two different kinds of dataset are investigated using state-of-the-art CNN models in this research work. One dataset(binary) has images of normal and tumor types, while another(multi-class) provides all images of tumors classified as glioma, meningioma, or pituitary. The experiments were conducted in these dataset with transfer learning from pre-trained weights from ImageNet as well as initializing the weights randomly. The experimental environment is equivalent for all models in this study in order to make a fair comparison. For both of the dataset, the validation set are same for all the models where train data is 60\% while the rest is 40\% for validation. With the proposed techniques in this research, the EfficientNet-B5 architecture outperforms all the state-of-the-art models in the binary-classification dataset with the accuracy of 99.75\% and 98.61\% accuracy for the multi-class dataset. This research also demonstrates the behaviour of convergence of validation loss in different weight initialization techniques.
\end{abstract}

\section{Introduction}
The brain tumor is one of the deadliest kinds of disease which is caused for the abnormal growth of abnormal cells that have formed in the brain. Some brain tumors are cancerous (malignant), while others are not (non-malignant). Since the brain is the control 
centre of the human body, developing tumors can put pressure 
on the skull and cause negative human health. The number of deaths due to brain tumors are increasing day by day. Early diagnosis is important for all brain tumors, as 
in all diseases. Early diagnosis of brain tumors is often made by 
magnetic resonance imaging (MRI) \cite{KUMAR2017510}.

Appropriate diagnosis of a brain tumor is critical for appropriate treatment planning and patient care. Manual classification of brain tumor MR images with similar structures or appearances is a complex and time-consuming task that is contingent on the radiologist's availability and experience in identifying and classifying the brain tumor. The radiologist classifies brain tumors in two ways: (i) Determine if the brain MR pictures are normal or abnormal. (ii) Classify aberrant brain magnetic resonance pictures into distinct tumor categories. Such manual classification of brain tumors is impracticable, non-reproducible, and time-consuming when dealing with vast amounts of MRI data. To address this issue, the automatic classification may be used to classify MR images of brain tumors with minimal interaction from radiologists.

The ability to make an accurate and early diagnosis of brain tumors is important for successful treatment. Because of the stage at which a tumor was discovered, its pathological type, and its severity, the treatment modality that is selected is dictated by these factors. Neuro-oncologists have benefited from the use of computer-aided diagnostic (CAD) techniques in a variety of ways. In neuro-oncology, CAD applications include tumor identification, classification, and grading. Brain tumor classification using CAD is a well-researched area.\cite{MOHAN2018139}. Grading glioma, a significant subtype of malignant tumor, is another area of research in this arena.\cite{wong2019pathology}.

Classification approaches based on traditional machine learning typically involve many steps, including preprocessing, feature extraction, feature selection, dimension reduction, and classification. Typically, feature extraction is dependent on the expert's topic expertise. It is difficult for a non-expert to do research using typical machine learning approaches. Traditional machine learning relies heavily on feature extraction, and classification accuracy is highly dependent on the extracted features. Two types of feature extraction exist. The first category includes low-level (global) characteristics such as intensity and texture. First- and second-order statistics (e.g., mean, standard deviation, and skewness) were obtained from the grey level co-occurrence matrix (GLMC), shape, wavelet transform, and Gabor feature.

In this study, Several CNN models have experimented with the same setup for both in the case of transfer learning and random weight initialization. The models that are experimented with are ResNet-50, ResNet18~\cite{he2016deep}, RexNext50~\cite{xie2017aggregated}, GoogleNet~\cite{szegedy2015going}, and EfficientNet~\cite{tan2019efficientnet}.
    
\section{Related Works}
It is essential to recognize and examine tumors at their early stage because doing so reduces the likelihood of becoming a victim. In order to determine the closeness of tumor prediction and categorization, many analysts directed numerous examinations employing a multitude of models. Various methods based on Deep Learning and machine learning have been developed over the last few years. As opposed to conventional classification approaches, the deep learning method does not rely on handcrafted feature extraction to achieve classification results. Deep learning approaches automate the learning of features from sample data that are difficult to understand and remember.
Using manually delineated tumor borders to extract features from the region of interest of T-1 MRI images, Cheng et al.\cite{cheng2015enhanced} proposed an approach where the best performance was achieved by Support-Vector-Machine model on bag of words (BOW) features. This is also considered to be one of the first works on the figshare brain MRI image dataset\cite{deepak2019brain}. For feature extraction, Ismail and Abdel-Qadar \cite{ismael2018brain} proposed the use of the Gabor filter and discrete wavelet transform, followed by the use of a multi-layer perceptron for classification. The shortcoming of these technologies is that they both rely on manual processes for feature extraction, which makes them less effective. CNNs have a distinct advantage in this situation because they do not require manually segmented regions and can extract all of the necessary features on their own. The drawback of these technologies is that they both rely on manual processes for feature extraction, which makes them less effective. CNNs have a distinct advantage in this situation because they do not require manually segmented regions and can extract all of the necessary features on their own. 
As a result, deep learning models based on Convolutional Neural Networks (CNNs) are being used by researchers to develop very efficient brain tumor classification algorithms. One drawback of CNN is that it takes a huge amount of data for training, which makes it difficult to use. Because of the complexity of brain tumors, a comparably deep convolutional neural network is required for classification from MRI images; yet, brain MRI image datasets are typically not very large. Due to these two diametrically opposed circumstances, a quandary is created: Transfer learning \cite{tan2018survey} is a fantastic approach for resolving this conundrum. It is possible to apply a deep pre-trained CNN model that was originally constructed for another similar application \cite{deepak2019brain} through transfer learning techniques. Khan Swati et al.\cite{swati2019brain} employed a pre-trained VGG- 19 model for diagnosing brain cancers from a figshare brain MRI image dataset, utilizing the notion of transfer learning to accomplish this. S.Deepak et al.\cite{deepak2019brain} used the same approach of transfer learning to apply a modified GoogLeNet model to the same figshare dataset~\cite{cheng_2017}.

Using enormous volumes of data to train on, deep learning \cite{smirnov2014comparison} demonstrates excellent performance and generalizability. This achievement may be attributed mostly to the rapid advancement in computer capacity, particularly through the use of graphics processing units, which allowed for the rapid creation of complicated deep learning algorithms. Deep learning architectures of many types have been created for a variety of applications, including computer vision classification, speech recognition, and object detection.
\section{Dataset}
Two different dataset are used in this research work. The binary classification dataset~\cite{hamada2021} has a total of 3000 samples. The multi-class dataset~\cite{cheng_2017} has 3064 samples.
\subsection{Dataset-1:Binary Classification}
\begin{figure}[H]
    \centering
   {\includegraphics[scale=.40]{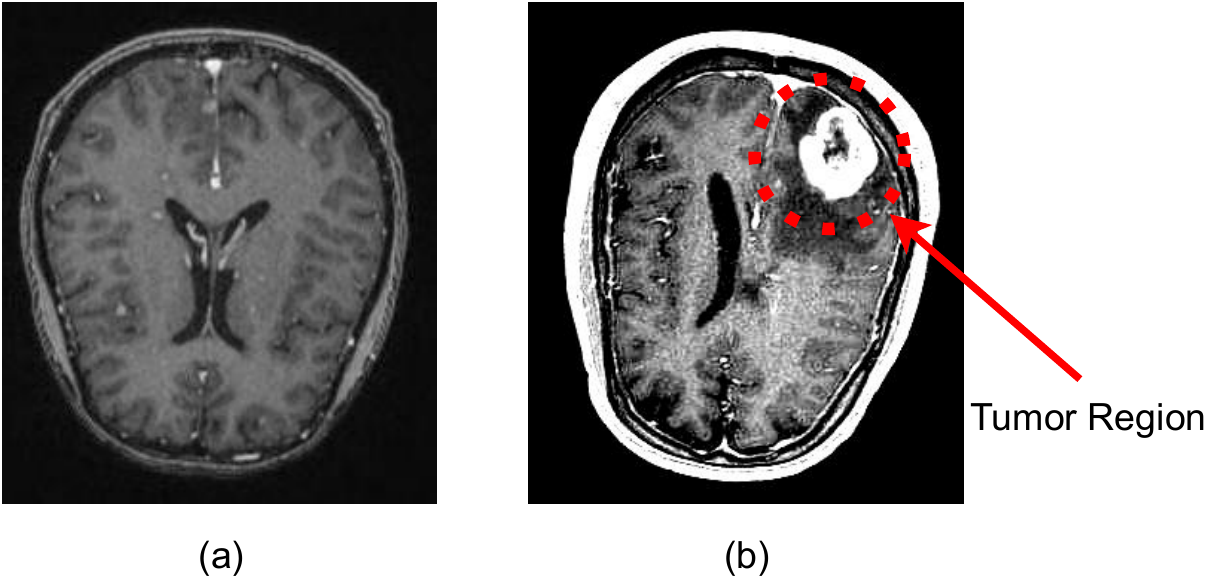}}
    \caption{Two sample RGB images from binary classification dataset. In (a), an RGB image is shown, where no region is can be detected as brain tumor. In (b), an RGB image with the specified brain tumor region is visualized.}%
    \label{fig:bin_dataset}%
\end{figure}

\begin{figure}[H]
    \centering
   {\includegraphics[scale=.40]{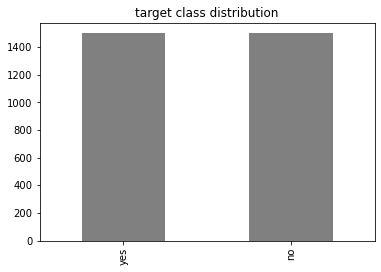} }
    \caption{Target class distribution for the binary-classification dataset.}%
    \label{fig:bin_dist}%
\end{figure}

The binary classification dataset, is mainly for detecting if there is a possibility of a brain tumor in the corresponding image or the image is normal. Here in~\ref{fig:bin_dataset}, samples from different classes are shown.
This research work conducted on this dataset has no class imbalance issue. The class distribution is shown in ~\ref{fig:bin_dist}.

\subsection{Dataset-2:Multiclass Classification}
\begin{figure}[H]
    \centering
   {\includegraphics[scale=.40]{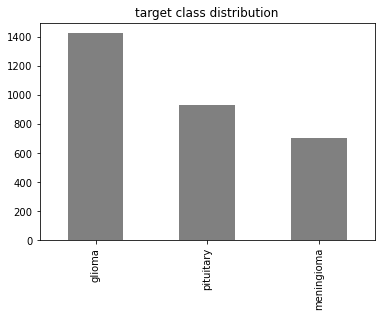} }
    \caption{Target class distribution for the multi-class dataset}%
    \label{fig:mul_dist}%
\end{figure}

This brain tumor dataset contains 3064 T1-weighted contrast-inhanced images
from 233 patients with three kinds of brain tumor: meningioma (708 slices), 
glioma (1426 slices), and pituitary tumor (930 slices). The class distribution is depicted in~\ref{fig:mul_dist}.
\begin{figure}[H]
    \centering
   {\includegraphics[scale=.40]{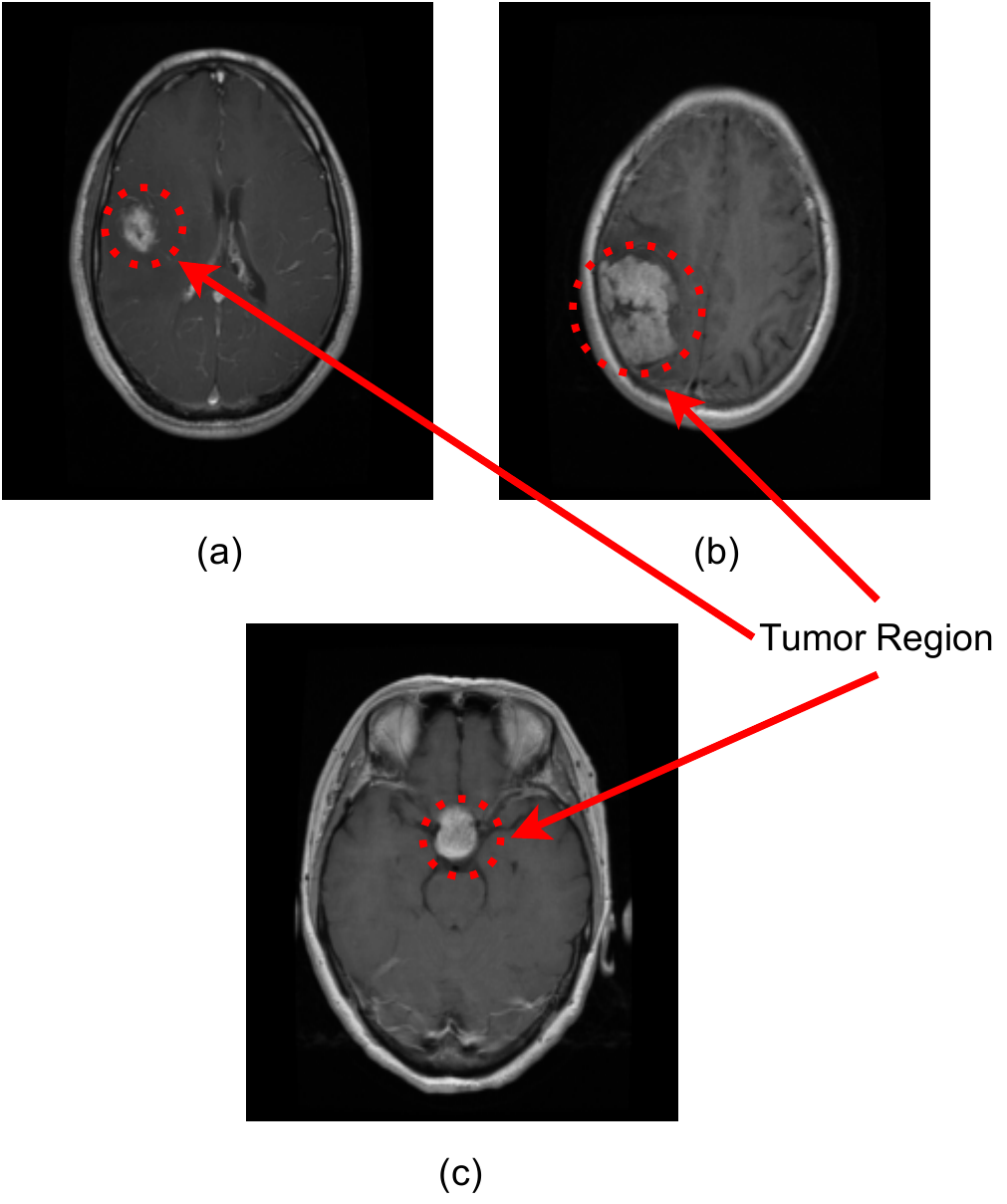} }
    \caption{Samples from each of the classes from the multi-class dataset. (a) Glioma, (b) Meningioma, (c) Pituitary.}%
    \label{fig:mul_dataset}%
\end{figure}
Demonstration of the tumor region in each of the class with samples is visualized in~\ref{fig:mul_dataset}. Here this image shows the different regions of the existence of the tumor.
\section{Performance Metrics}
All the performance metrics were kept the same during conduction of the experiments. The loss is calculated using~\ref{eqn:loss}.
\begin{equation}
\text { Cross-Entropy Loss }=-\sum_{i=1}^{N} y_{i} \times \log \hat{y}_{i}
\label{eqn:loss}
\end{equation}
In~\ref{eqn:acc}, the formula for calculating accuracy is shown. The other formula for calculating recall, precision and F1-score are shown in~\ref{eqn:recall},~\ref{eqn:precision}~\ref{eqn:f1} respectively.
\begin{equation}
\text { Acc. }=\frac{ \mathrm{T_P}+\mathrm{T_N}}{ \mathrm{T_P}+\mathrm{F_P}+\mathrm{F_N}+\mathrm{T_N}}
\label{eqn:acc}
\end{equation}
\begin{equation}
\text { Recall }=\frac{ \mathrm{T_P}}{ \mathrm{T_P}+\mathrm{F_N}}
\label{eqn:recall}
\end{equation}
\begin{equation}
\text { Precision }=\frac{ \mathrm{T_P}}{ \mathrm{T_P}+\mathrm{F_P}}
\label{eqn:precision}
\end{equation}
\begin{equation}
F 1=\frac{2 \times \text { precision } \times \text { recall }}{\text { precision }+\text { recall }}
\label{eqn:f1}
\end{equation}
The F1-Score is later calculated by precision and recall by the formula in~\ref{eqn:f1}.
Since the experiment is also conducted with the multi-class dataset. The precision, recall and F1-score are calculated using a macro average.
\section{Methodology}
\subsection{Data Preprocessing}
\subsubsection{Augmentation}
The augmentation step shown in ~\ref{fig:pipeline} is composed of rotation up-to ${ 20 \deg }$,
horizontal and vertical flip as augmentation techniques.  In the case of a vertical or horizontal flip, an image flip entails reversing the rows or columns of pixels. Flips are randomly chosen. In case, the input image, $\mathbf{X}\in \mathbb{R}^{m\times n}$, the horizontal and vertical flips are referred to by ~\ref{eqn:vert} and ~\ref{eqn:hor}.

\begin{equation}
        \mathbf{X}^h = \mathbf{X}_{i(n+1-j)}
        \label{eqn:vert}
\end{equation}
\begin{equation}
    \mathbf{X}^v = \mathbf{X}_{(m+1-i)j}
    \label{eqn:hor}
\end{equation}

\subsubsection{Channel-wise Standardization from ImageNet}
In this research, the channel-wise standardization is performed with~\ref{eqn:std}.
\begin{equation}
    \mathbf{X_o}=\frac{\mathbf{X_i}-\mu}{\delta}
    \label{eqn:std}
\end{equation}
In~\ref{eqn:std}, $\mathbf{X}_o$ represents each of the three channels, here $\mu \in \{0.485, 0.456, 0.406\}$ represents the mean for each of the channel. and $\delta\in\{0.229, 0.224, 0.225\}$ is the standard-deviation for each of the channel. The values for the set of $\mu$ and $\delta$ are taken from the imagenet data pre-processing formation. Since all the models have experimented with the pre-trained weights on imagenet~\cite{imagenet_cvpr09} in this setting.
\subsubsection{Normalization}
All the images are resized for $224\times 224$ a 2-dimensional tensor. The input images are then normalized using~\ref{eqn:normalization}.

\begin{equation}
    \hat{\boldsymbol{X}}=\frac{\boldsymbol{X}}{255.0}
    \label{eqn:normalization}
\end{equation}
Here, $\boldsymbol{X}$ is RGB resized input image matrix and $\boldsymbol{X}\in[0,255.0]$. And $\hat{\boldsymbol{X}}\in[0,1.0]$, which is normalized and directly processed by the deep learning methods.
\subsection{Proposed pipeline}
\begin{figure*}[htbp]
    \centering
   {\includegraphics[scale=.45]{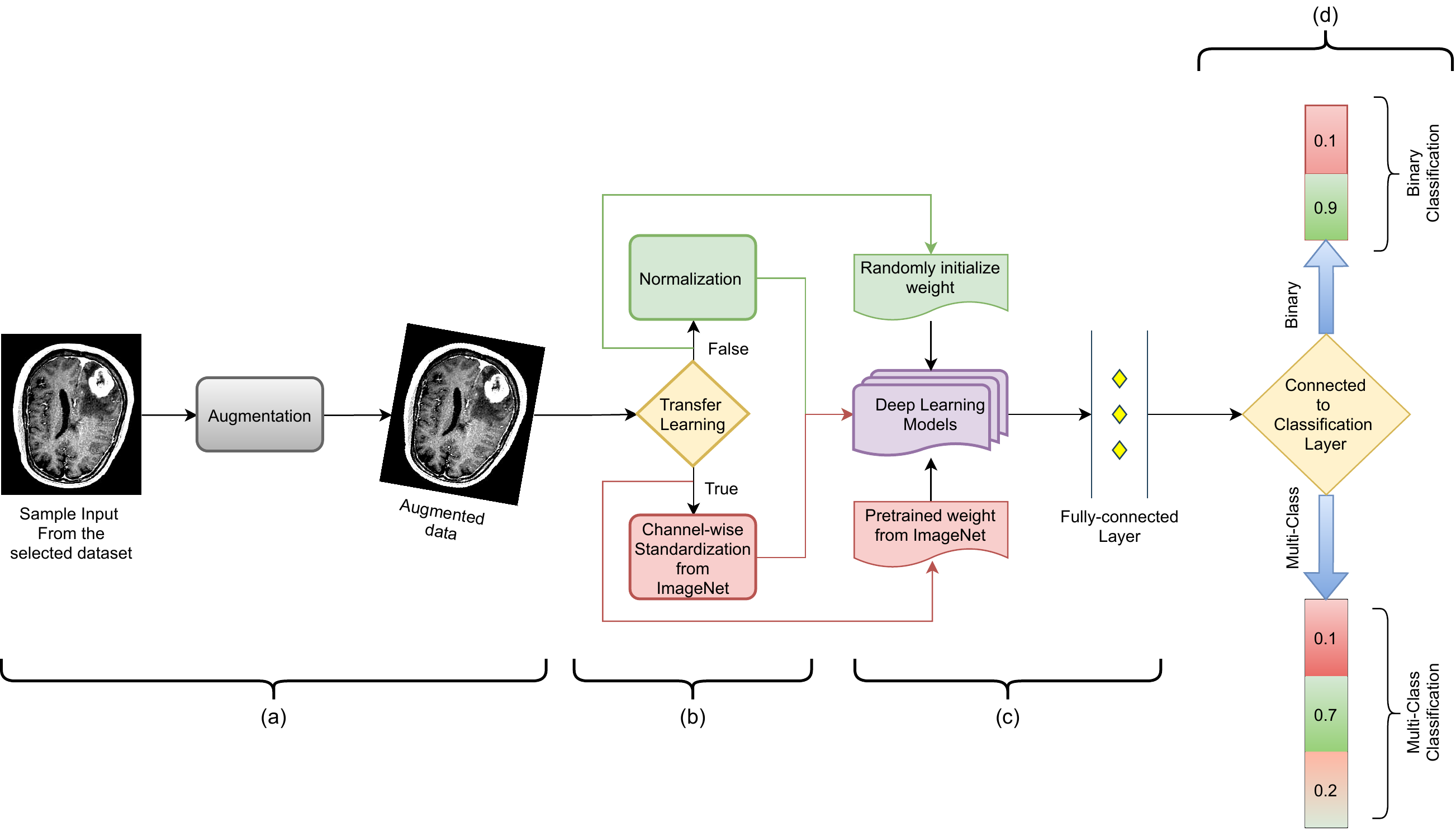} }
    \caption{Overview of the process. (a) The input image and augmentation process. (b) Normalization or Standardization according to the given scenario. (c) Weight initialization of the deep models based on the transfer-learning setting. (d) Modifying the classification layer according to the dataset.}%
    \label{fig:pipeline}%
\end{figure*}

The proposed pipeline of this research work is visualized in~\ref{fig:pipeline}. The algorithm that is used for training and evaluation is shown in~\ref{algo:training_and_evaluation}. The deep-learning models' weights are updated according to~\ref{eqn:weight_update}.
\begin{equation}
    \hat{\theta_i}=\theta_i-\alpha\left(\frac{\partial \tau_\ell }{\partial \theta_i}\right)
    \label{eqn:weight_update}
\end{equation}
Here, $\hat{\theta_i}$ represents the new weight of the $i^{th}$ layer. $\alpha$ represents the learning rate. $\theta_i$ is respective to the old weights, and $\frac{\partial \tau_\ell }{\partial \theta_i}$ represents the derivative of the total loss with respective to the weights. The  $\tau_\ell$ can be reffered to the loss from~\ref{eqn:loss}.

\begin{algorithm}

\DontPrintSemicolon
\KwIn{ Training dataset $T$, Validation dataset $V$, model $M_{\theta_i}$, no. of epochs $N$}
\KwResult{Optimized model $\hat{M_{\theta_i}}$}

Randomly initialize model parameters $\theta$\;
$l_v \leftarrow \infty$\;
\For{$i \leftarrow 1\ to\ N$}{
$M_{\theta_i} \leftarrow T$\;
Calculate training loss $L_T$ according to eqn.~\ref{eqn:loss} \;
Update model parameters $\theta$ of $M_{\theta_i}$ according to eqn.~\ref{eqn:weight_update}\;
$M_{\theta_i} \leftarrow V$\;
Calculate validation loss $L_V$ according to equation~\ref{eqn:loss}\;

$\hat{M_{\theta_i}} \leftarrow M_{\theta_i}$, $l_v \leftarrow L_v$,   If $L_v \leqslant l_v,\forall_i \in \{1, . . . , i\}$\;
}

Save and return $\hat{M_{\theta_i}}$ \;
\caption{Model Parameters Optimization\label{algo:training_and_evaluation}}
\end{algorithm}

~\ref{algo:training_and_evaluation} saves the best model from the minimum validation loss achieved strategy. All the models experimented with are saved in the same strategy. This technique also reduces the overfitting to the unseen dataset.
\subsection{Experimental Setup}
\begin{figure}[H]
    \centering
    \includegraphics[scale=.50]{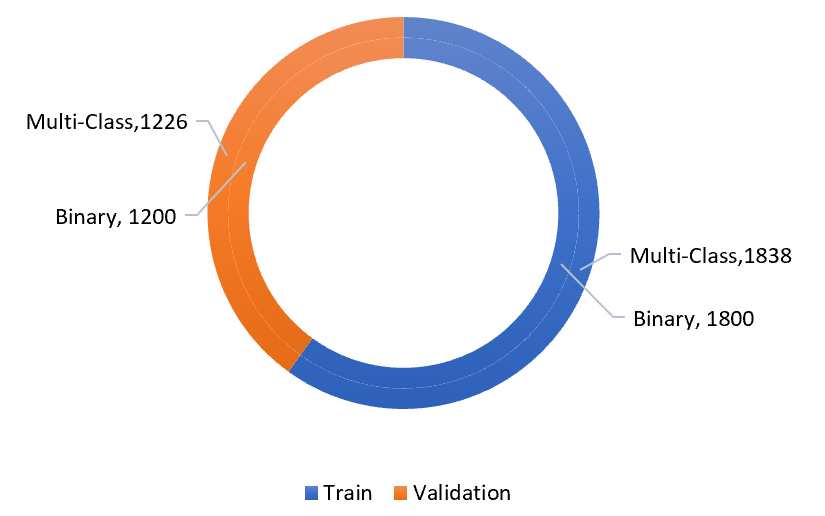}
    \caption{Train-Validation Split of the two dataset.}
    \label{fig:split}
\end{figure}

\begin{figure}[H]
    \centering
    \includegraphics[scale=.48]{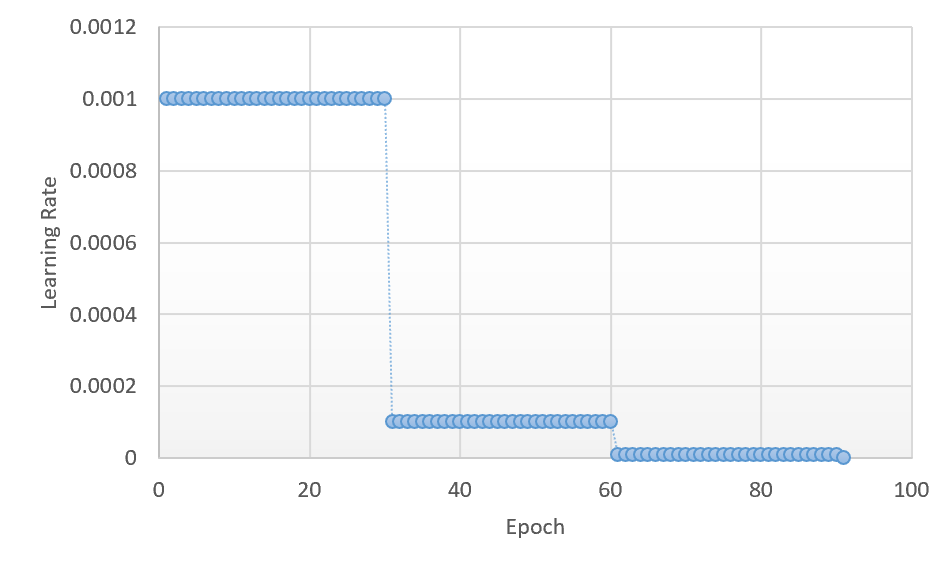}
    \caption{Effect of the scheduler in learning rate in different epochs}
    \label{fig:learning_rate}
\end{figure}

Both of the dataset in this research are separated into two groups: train and validation. To ensure the correct capability of the models for unseen data validation sets were the same for all the models. The split of the dataset is shown in~\ref{fig:split}.
From~\ref{fig:split}, it is visible that, 60\% of data are used for training while the remaining 40\% are for validation. This way by keeping the validation(i.e: unseen samples) set amount close to the training set amount to ensure the robustness of the models for unseen data. To optimize the model, \textbf{\textit{Adam}} optimizer with \textbf{\textit{StepLR}} scheduler is used. The learning rate and the effect of this scheduler is visualized in~\ref{fig:learning_rate}.
In~\ref{fig:learning_rate}, it is visible that the initial learning rate was 0.001 and the step size is 30. Due to this fact, the learning rate is decremented by 0.1 in every $30^{th}$ multiple. All the models are trained with 100 epochs.

\section{Result Analysis}

\subsection{Binary-Classification dataset}
The performance result for the validation set for different deep models is shown in this section. When the models are trained by initializing the weights randomly, it is observable from~\ref{tab:bin} that EfficientNet-B7 achieved the best result among all the state-of-the-art methodologies.

\begin{table*}
\centering
\caption{Binary Classification Without Transfer Learning}
\label{tab:bin}
\begin{tabular}{cccccc} 
\hline
\rowcolor[rgb]{0.753,0.753,0.753} Model Name & Accuracy          & F1-Score          & Precision         & Recall            & Loss               \\
ResNet-50                                    & 0.988333          & 0.986491          & 0.988056          & 0.989270          & 0.057198           \\
ResNet18                                     & 0.990833          & 0.989153          & 0.990333          & 0.991603          & 0.038533           \\
ResNext50                                    & 0.990833          & 0.989713          & 0.990778          & 0.991802          & 0.045017           \\
GoogleNet                                    & 0.990000          & 0.989226          & 0.990944          & 0.990389          & 0.041164           \\
EfficientNet B0                              & 0.987500          & 0.986165          & 0.988690          & 0.987857          & 0.040588           \\
EfficientNet B1                              & 0.990833          & 0.989826          & 0.992000          & 0.990333          & 0.052381           \\
EfficientNet B2                              & 0.990000          & 0.988828          & 0.990222          & 0.990468          & 0.051303           \\
EfficientNet B3                              & 0.989167          & 0.988282          & 0.990389          & 0.989444          & 0.045488           \\
EfficientNet B4                              & 0.989167          & 0.987735          & 0.989222          & 0.989635          & 0.038901           \\
EfficientNet B5                              & 0.990000          & 0.988061          & 0.989389          & 0.990325          & 0.048566           \\
EfficientNet B6                              & 0.989167          & 0.987981          & 0.989746          & 0.989802          & 0.039184           \\
\textbf{EfficientNet B7}                     & \textbf{0.995833} & \textbf{0.995000} & \textbf{0.995167} & \textbf{0.996357} & \textbf{0.018193}  \\
\hline
\end{tabular}
\end{table*}
In~\ref{tab:bin_imgnet}, the results for the models are shown, here the models' weights are initialized from ImageNet pretrained weights. It is observable from~\ref{tab:bin} and ~\ref{tab:bin_imgnet}, EfficientNet family outperforms all the state-of-the-art models. However, with transfer-learning Efficient-B5 can outperform the B7 version. 

   

\begin{figure}[H]
    \centering
    \subfloat[With-Out Transfer Learning]{{\includegraphics[scale=.17]{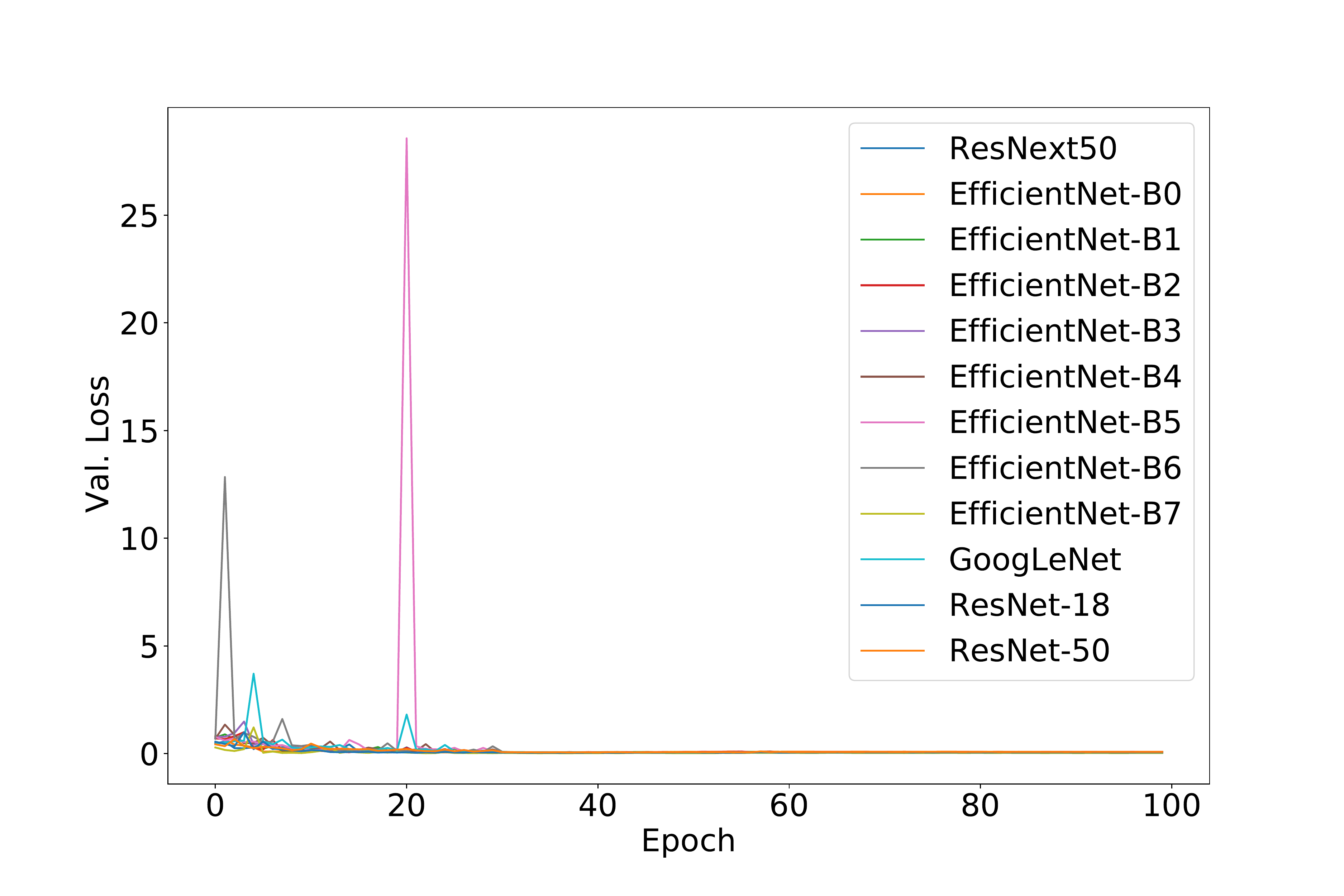} }}%
    \qquad
    \subfloat[With Transfer Learning]{{\includegraphics[scale=.17]{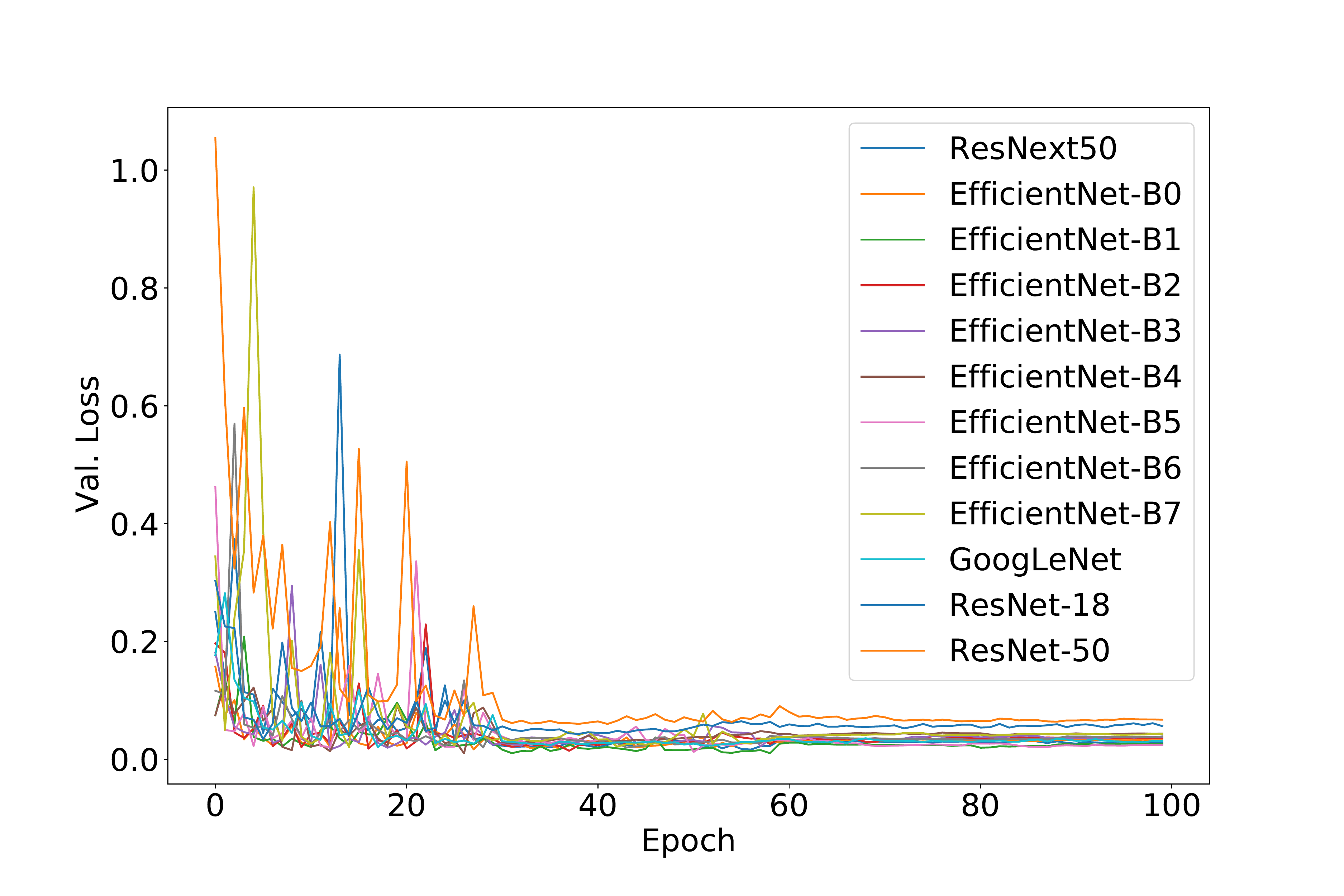} }}%
    \caption{Epoch vs. validation loss for the binary classification dataset}%
    \label{fig:bin_loss}%
\end{figure}

\begin{table*}
\centering
\caption{Binary Classification With Transfer Learning}
\label{tab:bin_imgnet}
\begin{tabular}{cccccc} 
\hline
\rowcolor[rgb]{0.753,0.753,0.753} Model Name & Accuracy          & F1-Score          & Precision         & Recall            & Loss               \\
ResNet-50                                    & 0.990833          & 0.989277          & 0.990278          & 0.991437          & 0.060031           \\
ResNet18                                     & 0.994167          & 0.993184          & 0.994500          & 0.994413          & 0.034172           \\
ResNext50                                    & 0.995833          & 0.995123          & 0.995444          & 0.996246          & 0.016570           \\
GoogleNet                                    & 0.995000          & 0.992858          & 0.993389          & 0.994111          & 0.020670           \\
EfficientNet B0                              & 0.995833          & 0.995521          & 0.996722          & 0.995690          & 0.016866           \\
EfficientNet B1                              & 0.996667          & 0.995123          & 0.995611          & 0.996357          & 0.010314           \\
EfficientNet B2                              & 0.995833          & 0.995123          & 0.995889          & 0.996079          & 0.014472           \\
EfficientNet B3                              & 0.996667          & 0.996491          & 0.997111          & 0.996722          & 0.016532           \\
EfficientNet B4                              & 0.996667          & 0.996367          & 0.997389          & 0.996278          & \textbf{0.010307}  \\
\textbf{EfficientNet B5}                     & \textbf{0.997500} & \textbf{0.997460} & \textbf{0.997667} & \textbf{0.997833} & 0.012521           \\
EfficientNet B6                              & 0.995833          & 0.995521          & 0.996556          & 0.995667          & 0.019978           \\
EfficientNet B7                              & 0.996667          & 0.996244          & 0.996944          & 0.996556          & 0.020426           \\
\hline
\end{tabular}
\end{table*}
The validation loss over the entire training and evaluation duration is shown in~\ref{fig:bin_loss}.

From~\ref{fig:bin_loss} it is observable that the convergence of the models happens earlier and in a better manner when they are not pre-trained. However, with the transfer-learning mechanism, they can reach a better minimum loss value hence improved performance.
\subsection{Multi-Class Classification dataset}
In the case of multi-class classification dataset, the scenario is different. Though the training, validation and model selection are all same, here in this~\ref{tab:mult} EfficientNet-B3 achieves the best performance when training the models without transfer learning. However, while the models are pre-trained, Efficient-B5 achieves the best performance which also outperforms all the models when they were trained by initializing the weights randomly.


\begin{table*}
\centering
\caption{Multi-class Classification Without Transfer Learning}
\label{tab:mult}
\begin{tabular}{cccccc} 
\hline
\rowcolor[rgb]{0.753,0.753,0.753} Model Name & Accuracy          & F1-Score          & Precision         & Recall            & Loss               \\
ResNet-50                                    & 0.965742          & 0.950289          & 0.956389          & 0.956335          & 0.135912           \\
ResNet18                                     & 0.967374          & 0.954463          & 0.959978          & 0.961796          & 0.127369           \\
ResNext50                                    & 0.974715          & \textbf{0.966937} & \textbf{0.971211} & \textbf{0.972684} & 0.104295           \\
GoogleNet                                    & 0.967374          & 0.955982          & 0.962534          & 0.962213          & 0.103805           \\
EfficientNet B0                              & 0.974715          & 0.962782          & 0.968303          & 0.968109          & 0.101448           \\
EfficientNet B1                              & 0.964927          & 0.950665          & 0.958329          & 0.956521          & 0.117937           \\
EfficientNet B2                              & 0.973899          & 0.959503          & 0.964196          & 0.964741          & 0.097664           \\
\textbf{EfficientNet B3}                     & \textbf{0.977977} & 0.963085          & 0.969989          & 0.966128          & 0.097950           \\
EfficientNet B4                              & 0.970636          & 0.958349          & 0.965131          & 0.962987          & \textbf{0.092380}  \\
EfficientNet B5                              & 0.968189          & 0.955456          & 0.962878          & 0.959603          & 0.138125           \\
EfficientNet B6                              & 0.968189          & 0.948156          & 0.953484          & 0.953101          & 0.117775           \\
EfficientNet B7                              & 0.973899          & 0.960490          & 0.967995          & 0.966387          & 0.105291           \\
\hline
\end{tabular}
\end{table*}
However, in terms of other metrics such as F1-Score, Precision and Recall, ResNext50 achieves the best performance when training the model without transfer learning.

\begin{table*}
\centering
\caption{Multiclass Classification With Transfer Learning}
\label{tab:mult_imgnet}
\begin{tabular}{cccccc} 
\hline
\rowcolor[rgb]{0.753,0.753,0.753} Model Name & Accuracy          & F1-Score          & Precision         & Recall            & Loss               \\
ResNet-50                                    & 0.970636          & 0.961424          & 0.966418          & 0.966467          & 0.119400           \\
ResNet18                                     & 0.973083          & 0.962029          & 0.969005          & 0.966232          & 0.121884           \\
ResNext50                                    & 0.972268          & 0.958944          & 0.964999          & 0.963014          & 0.112022           \\
GoogleNet                                    & 0.968189          & 0.954450          & 0.961682          & 0.962770          & 0.126151           \\
EfficientNet B0                              & 0.986134          & 0.981124          & 0.984231          & 0.983071          & 0.093312           \\
EfficientNet B1                              & 0.982055          & 0.974132          & 0.980895          & 0.976115          & 0.078309           \\
EfficientNet B2                              & 0.977977          & 0.967653          & 0.973386          & 0.972309          & 0.074401           \\
EfficientNet B3                              & 0.982055          & 0.974204          & 0.979373          & 0.976799          & 0.078525           \\
EfficientNet B4                              & 0.986134          & \textbf{0.981355} & \textbf{0.986333} & \textbf{0.983324} & 0.081955           \\
\textbf{EfficientNet B5}                     & \textbf{0.986134} & 0.979857          & 0.984702          & 0.981408          & 0.084033           \\
EfficientNet B6                              & 0.985318          & 0.978629          & 0.982889          & 0.980864          & \textbf{0.073885}  \\
EfficientNet B7                              & 0.980424          & 0.970646          & 0.976001          & 0.972848          & 0.095240           \\
\hline
\end{tabular}
\end{table*}
In the case of pre-trained models, the EfficientNet B4 version achieves the best performance for F1-Score, Precision and recall.

\begin{table*}
\centering
\caption{Comparison with related works using Figshare(Multi-Class) dataset}
\resizebox{\linewidth}{!}{%
\begin{tabular}{|l|l|l|l|l|l|l|} 
\hline
\begin{tabular}[c]{@{}l@{}}\\Work\end{tabular} & Method                                    & Accuracy             & F1-Score             & Precision            & Recall               & Loss                  \\ 
\hline
Swati et al. (2019)                            & Fine tuned VGG-19 \cite{swati2019brain}                        & 94.8                 & -                    & -                    & -                    & -                     \\ 
\hline
S. Deepak et al.(2019)                         & deep CNN-SVM \cite{deepak2019brain}                             & 97.1                 & -                    & -                    & -                    & -                     \\ 
\hline
S. Deepak et al.(2021)                         & SNN—k-NN \cite{deepak2021brain}                                 & 92.6                 & -                    & -                    & -                    & -                     \\ 
\hline
\textbf{This Research Best Result}                             & \textbf{EffNet with Transfer-Learned CNN} & \textbf{98.613}                & \textbf{98.135}                & \textbf{98.633}                & \textbf{98.33}                & \textbf{0.0738}                 \\ 
\hline
\multicolumn{1}{l}{}                           & \multicolumn{1}{l}{}                      & \multicolumn{1}{l}{} & \multicolumn{1}{l}{} & \multicolumn{1}{l}{} & \multicolumn{1}{l}{} & \multicolumn{1}{l}{} 
\end{tabular}
}
\end{table*}

\begin{table*}[]
\caption{Comparison with related works using Br35H(Binary) dataset}
\resizebox{\linewidth}{!}{%
\begin{tabular}{|l|l|l|l|l|l|l|}
\hline
\begin{tabular}[c]{@{}l@{}}\\Work\end{tabular} & Method            & Accuracy & F1-Score & Precision & Recall & Loss \\
\hline
Jaeyong Kang et al. (2021)    & K-NN \cite{kang2021mri} &  98.17     & -        & -         & -      & -    \\
\hline
S. Deepak et al.(2019) & deep CNN-SVM  \cite{deepak2021brain}    & 97.1     & -        & -         & -      & -    \\
\hline
S. Deepak et al.(2021) & SNN—k-NN   \cite{deepak2021brain}       & 92.6     & -        & -         & -      & - \\
\hline
\textbf{This Research Best Result} & \textbf{EffNet with Transfer-Learned CNN} &\textbf{99.75}     & \textbf{99.74}        & \textbf{99.76}         & \textbf{99.78} & \textbf{0.010} \\
\hline
\multicolumn{1}{l}{}                           & \multicolumn{1}{l}{}                      & \multicolumn{1}{l}{} & \multicolumn{1}{l}{} & \multicolumn{1}{l}{} & \multicolumn{1}{l}{} & \multicolumn{1}{l}{}
\end{tabular}
}
\end{table*}

\begin{figure}[H]
    \centering
    \subfloat[With-Out Transfer Learning]{{\includegraphics[scale=.17]{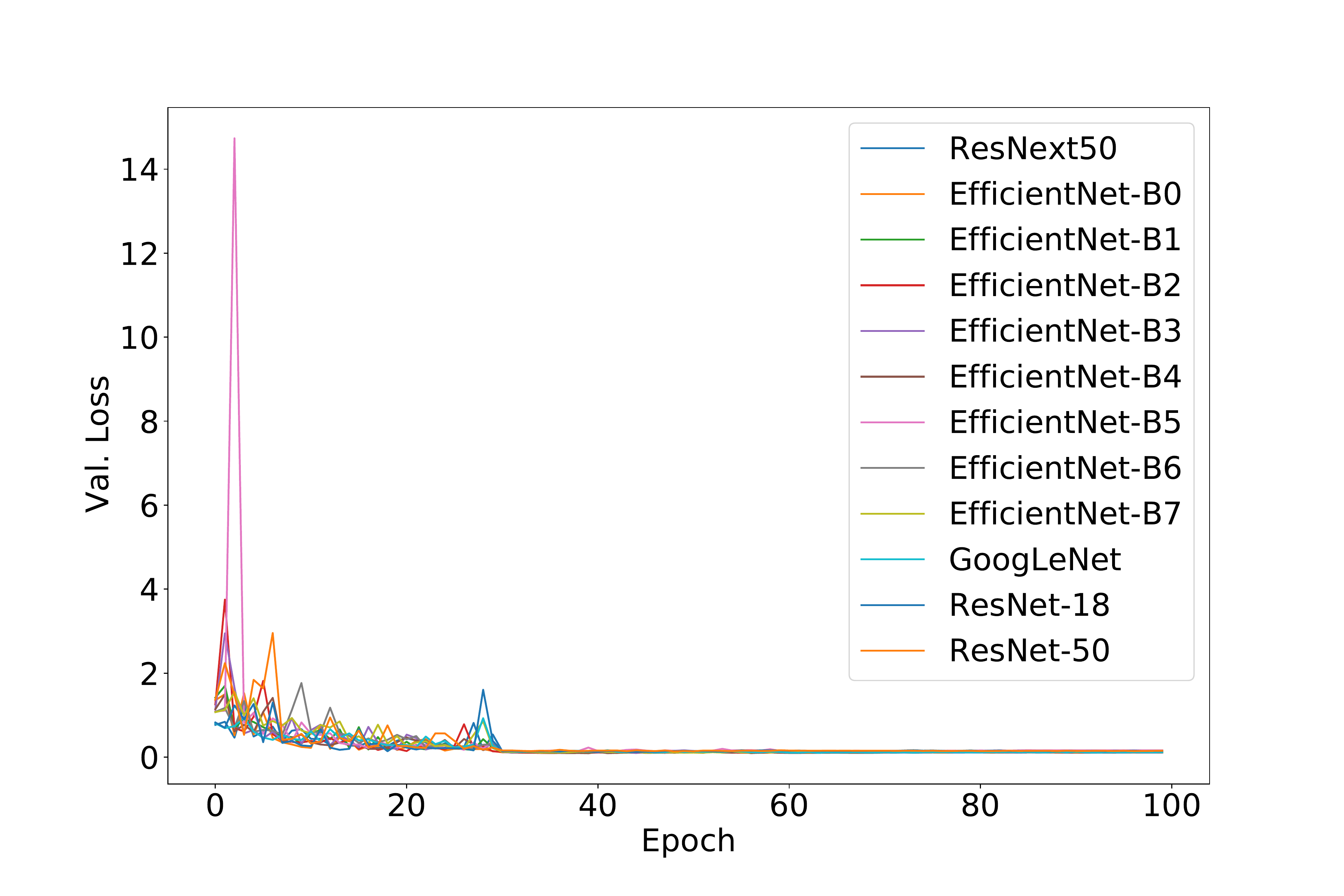} }}%
    \qquad
    \subfloat[With Transfer Learning]{{\includegraphics[scale=.17]{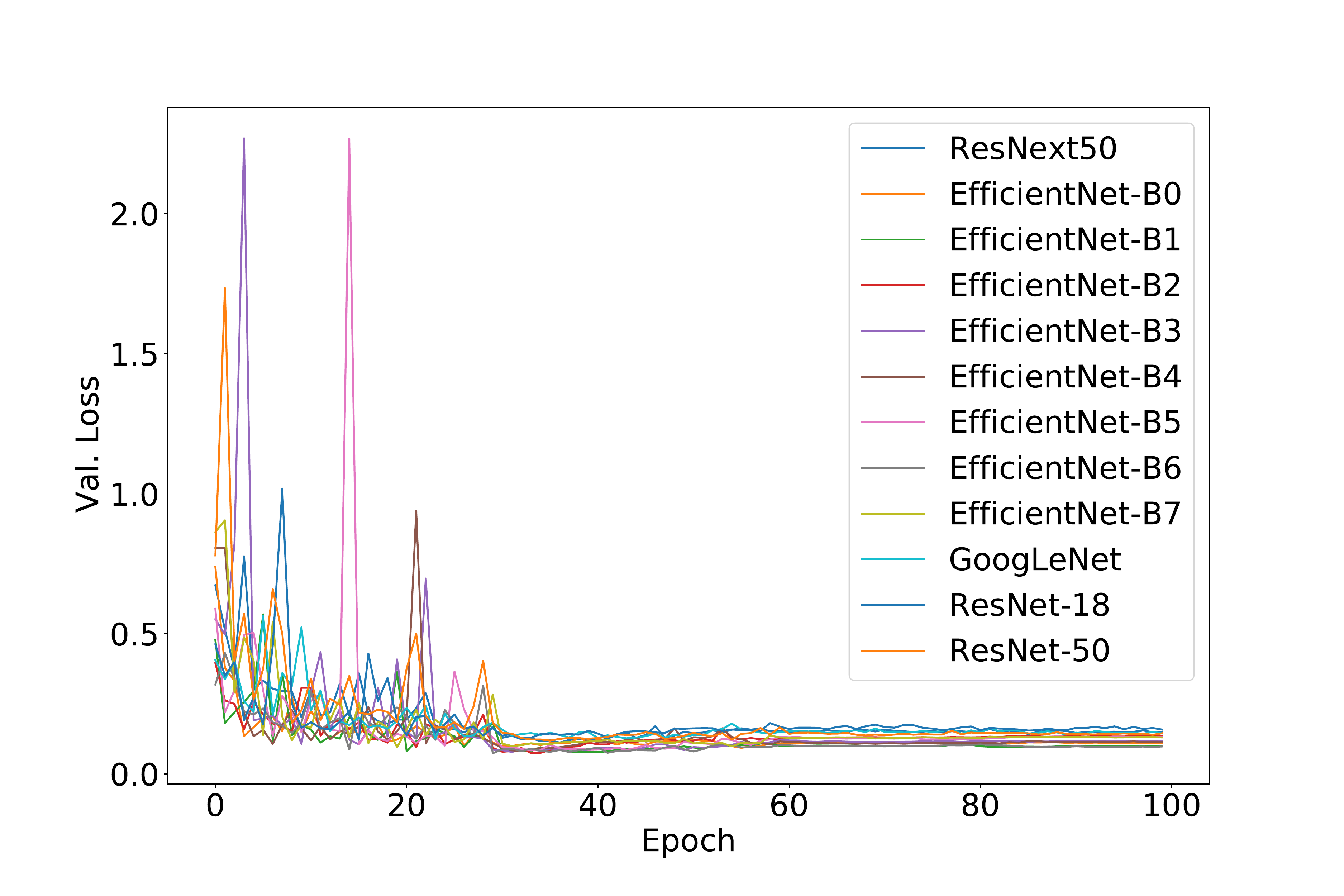} }}%
    \caption{Epoch vs. validation loss for the Multi-Class classification dataset}%
    \label{fig:multi_loss}%
\end{figure}
The situation for validation loss curve visualized in~\ref{fig:multi_loss} for the multi-class dataset is similar as the binary classification dataset. Here in this case also has an earlier convergence while training the models from scratch, they reach a better minimum when they are trained from pre-trained weights.
\section{Conclusion}
In this paper, findings of the behaviour of the deep CNN models are demonstrated. The environment for the experiments was similar for the models to make a fair comparison among the models. From the experiments, it is found that the EfficientNet architectures can achieve the best result for brain tumor classification datasets. In addition to this, the convergence of the models with respect to different settings is also demonstrated. In spite of having an earlier convergence when training the models from the random weight initialization technique, the models can reach a better minimum loss when they are trained from pre-trained weights.

%
%
%

\bibliographystyle{splncs04}
\bibliography{1.main}

\end{document}